\documentclass[fleqn,usenatbib]{mnras}

\usepackage{newtxtext,newtxmath}

\usepackage[T1]{fontenc}

\DeclareRobustCommand{\VAN}[3]{#2}
\let\VANthebibliography\thebibliography
\def\thebibliography{\DeclareRobustCommand{\VAN}[3]{##3}\VANthebibliography}

\usepackage{graphicx}	
\usepackage{amsmath}	
\usepackage{verbatim}   
\usepackage{scalerel}
\usepackage{tikz}
\usepackage{soul}       

\soulregister\cite7
\soulregister\citet7
\soulregister\citep7
\soulregister\ref7
\soulregister\pageref7

\usetikzlibrary{svg.path}

\definecolor{orcidlogocol}{HTML}{A6CE39}
\tikzset{
  orcidlogo/.pic={
    \fill[orcidlogocol] svg{M256,128c0,70.7-57.3,128-128,128C57.3,256,0,198.7,0,128C0,57.3,57.3,0,128,0C198.7,0,256,57.3,256,128z};
    \fill[white] svg{M86.3,186.2H70.9V79.1h15.4v48.4V186.2z}
                 svg{M108.9,79.1h41.6c39.6,0,57,28.3,57,53.6c0,27.5-21.5,53.6-56.8,53.6h-41.8V79.1z M124.3,172.4h24.5c34.9,0,42.9-26.5,42.9-39.7c0-21.5-13.7-39.7-43.7-39.7h-23.7V172.4z}
                 svg{M88.7,56.8c0,5.5-4.5,10.1-10.1,10.1c-5.6,0-10.1-4.6-10.1-10.1c0-5.6,4.5-10.1,10.1-10.1C84.2,46.7,88.7,51.3,88.7,56.8z};
  }
}

\newcommand\orcidicon[1]{\href{https://orcid.org/#1}{\mbox{\scalerel*{
\begin{tikzpicture}[yscale=-1,transform shape]
\pic{orcidlogo};
\end{tikzpicture}
}{|}}}}

\title[Winds from spherically-stratified starburst nuclei]{Dynamics of hot galactic winds launched from spherically-stratified starburst cores}

\author[Nguyen et al.]{Dustin D.~Nguyen $^{1,2,3}$\thanks{E-mail: dnguyen.phys@gmail.com} \orcidicon{0000-0002-1875-6522}, Todd A.~Thompson $^{1,2,4}$ \orcidicon{0000-0003-2377-9574} , Evan E.~Schneider$^{5,6}$  \orcidicon{0000-0001-9735-7484}, \newauthor Sebastian Lopez $^{1,4}$\orcidicon{0000-0002-2644-0077}, Laura A. Lopez$^{1,4,7}$ \orcidicon{0000-0002-1790-3148}   \\
  $^{1}$Center for Cosmology and Astro-Particle Physics (CCAPP), Ohio State University, 191 W. Woodruff Ave, Columbus, OH 43210, USA \\
  $^{2}$Department of Physics, Ohio State University, 191 W. Woodruff Ave, Columbus, OH 43210, USA \\
  $^{3}$Computational Physics and Methods Group (CCS-2), Los Alamos National Laboratory, Los Alamos, NM 87545, USA \\
  $^{4}$Department of Astronomy, Ohio State University, 140 W.~18th Ave, Columbus, OH 43210, USA \\
  $^{5}$Department of Physics and Astronomy, University of Pittsburgh, Pittsburgh, PA 15260, USA \\ 
  $^{6}$Pittsburgh Particle Physics, Astrophysics, and Cosmology Center (PITT PACC), University of Pittsburgh, Pittsburgh, PA 15260, USA \\
  $^{7}$Flatiron Institute, Center for Computational Astrophysics, NY 10010, USA
  }

\date{Accepted XXX. Received YYY; in original form ZZZ}

\pubyear{2022}

\begin{document}
\label{firstpage}
\pagerange{\pageref{firstpage}--\pageref{lastpage}}
\maketitle

\begin{abstract}
 The analytic galactic wind model derived by Chevalier and Clegg in 1985 (CC85) assumes \textit{uniform} energy and mass-injection within the starburst galaxy nucleus. However, the structure of nuclear star clusters, bulges, and star-forming knots are non-uniform. We generalize to cases with spherically-symmetric energy/mass injection that scale as $r^{-\Delta}$ within the starburst volume $R$, providing solutions for $\Delta = 0$, 1/2, 1, 3/2, and 2. In marked contrast with the CC85 model ($\Delta=0$), which predicts zero velocity at the center, for a singular isothermal sphere profile ($\Delta=2$), we find that the flow maintains a \textit{constant} Mach number of $\mathcal{M}=\sqrt{3/5} \simeq 0.77$ throughout  the volume. The fast interior flow can be written as $v_{r < R} =  (\dot{E}_T/3\dot{M}_T)^{1/2} \simeq 0.41 \, v_\infty$, where $v_\infty$ is the asymptotic velocity, and $\dot{E}_T$ and $\dot{M}_T$ are the total energy and mass injection rates. For $v_\infty \simeq 2000 \, \mathrm{km \, s^{-1}}$, $v_{r<R} \simeq 820 \, \mathrm{km\, s^{-1}}$ throughout  the wind-driving region. The temperature and density profiles of the non-uniform models may be important for interpreting spatially-resolved maps of starburst nuclei. We compute velocity resolved spectra to contrast the $\Delta=0$ (CC85) and $\Delta=2$ models. Next generation X-ray space telescopes such as XRISM may assess these kinematic predictions.
\end{abstract}


\section{Introduction} 
Galactic winds are important to the process of galaxy formation and evolution \citep[see][]{Veilleux2005,Zhang2018,Veilleux2020}. They are commonly found in rapidly star-forming galaxies at both low and high redshift \citep{Martin2005,Rubin2014}, act to modulate star formation, shape the stellar mass and mass-metallicity relations \citep{Peeples2011,Ma2016}, and advect metals into the circumgalactic and intergalactic medium \citep{Borthakur2013,Werk2016}.

Galactic outflows are observed to be multi-phase. The hot, $T \geq 10^7\,$K, phase is observed in X-rays and is often compared to the CC85 wind model (e.g., \citealt{Strickland2009,Lopez2020,Lopez2022}). The CC85 model assumes uniform energy and mass-injection within a sphere of radius $R$ ($\sim 0.1 - 0.5\,\mathrm{kpc})$, which drives a flow that has the characteristic solution of transitioning from sub to supersonic at $R$. Outside of the sphere there are no energy and mass sources and the flow undergoes adiabatic expansion (i.e., $T\propto r^{-4/3}, \ \rho \propto r^{-2}, \ \mathrm{and} \ v \propto r^0$). 

There have been many modifications to CC85. These semi-analytic studies typically relax the assumption of an adiabatic wind by including additional physics such as radiative cooling, gravity, radiation pressure, non-equillibrium ionization, non-spherical flow geometries, and/or mass-loading of swept up material \citep[see][]{Wang1995,Suchkov1996,Silich2004,Thompson2016,Bustard2016,Yu2020,Nguyen2021,Fielding2022,Sarkar2022}. Other studies have numerically considered uniform wind-driving cylinders \citep{Strickland2000} and rings \citep{Nguyen2022}, and non-uniform injection within cold galactic disks \citep{Tanner2016,Schneider2020}. 

Star formation in nuclear star clusters is inherently non-uniform. Embedded stellar clusters display either multi-peaked surface density distributions or highly concentrated surface density distributions \citep{Lada2003}. Nuclear star clusters and bulges are observed to be compact and non-uniform \citep{Boker2002}. Consequently, a self-consistent wind model needs to consider non-uniform sources within the wind-driving region (WDR). 

In this work, in contrast with uniform injection, we consider spherically-symmetric volumetric energy and mass injection, $q\,\mathrm{[ergs\,cm^{-3}]}$ and $Q\,\mathrm{[g\,cm^{-3}]}$ respectively, that scales as $q \propto Q \propto r^{-\Delta}$ within the WDR. \citet{Zhang2014} present the solutions for arbitrary $\Delta$ models but do not present a study on the bulk gas dynamics and thermodynamics of these models. \citet{Silich2011} presents wind models with non-uniform mass and energy injection modeled with an exponential function as $q \propto Q \propto \exp(r/R)$. Both \citet{Palou2013} and \citet{Bustard2016} consider a Schuster distribution of sources that scale as $q \propto Q \propto (1-r^2/R^2)^\zeta$ (with the latter reference taking $\zeta=0$). In these previous works the sonic point shifts away from $R$, as $q$ and $Q$ are taken to be non-uniform. 

Here we calculate the structure of $r^{-\Delta}$ models for $r<R$. Similar to CC85, we assume that the supernovae efficiently thermalize their energy and drive a wind. We extend \citet{Zhang2014} by exploring how the kinematic and thermodynamic properties of the flow change over different injection slopes $\Delta$. The CC85 model ($\Delta=0$) predicts flat temperature, densities, and pressure within the WDR and zero velocity at the center which linearly accelerates to become supersonic at the starburst ridge. We find the non-uniform models produce flows that are denser and faster than the CC85 flows within $R$. Notably, for a model representative of a galactic density profile with a constant rotation curve, an isothermal sphere ($\rho_\mathrm{sources} \propto r^{-2}$), the outflow maintains $\mathcal{M} = \sqrt{3/5} \simeq 0.77$ flow throughout the WDR. We verify these results using 3D hydrodynamic simulations with the \texttt{Cholla} code for $\Delta =  1/2, \ 1, \ 3/2, \ \mathrm{and} \ 2$ models. We then focus on the observational characteristics of these non-uniform injection wind models, finding that the fast subsonic winds ($v_{r<R} \simeq 0.4 \, v_\infty$, see Eq.~\ref{eq:v_Constant_vasymp}), leads to horn-like features in resolved line profiles (Fig.~\ref{fig:surface_bright}) which may be observed by XRISM. 

In \autoref{sec:solutions}, we write down the hydrodynamic equations, derive the self-similar analytic Mach number, physical, dimensionless solutions, and take central limits of these solutions. In \autoref{sec:cholla}, we run 3D hydrodynamic simulations, confirm the derived analytics, and construct X-ray surface brightness, brightness vs. height profiles, and velocity resolved line profiles. In \autoref{sec:Summary}, we provide a synthesis of this work, discuss how the models predict outflow velocities that can be resolved by XRISM, how the different $T$ and $n$ profiles may be important in interpreting spatially-resolved maps for the interior of starburst superwinds, and consider future research directions. 
\section{Hydrodynamic Equations}
\label{sec:solutions}
In the absence of rotation, gravity, and radiative cooling, the hydrodynamic equations for a steady-state spherically expanding flow are \citep[see][]{Chevalier1985}:
\begin{align}
    & \frac{1}{r^2} \frac{d}{dr} ( \rho v r^2) = q, \label{eq:continuity} \\ 
    & v \frac{dv}{dr} = - \frac{1}{\rho} \frac{dP}{dr} - \frac{ q v }{\rho}, \label{eq:momentum} \\ 
    & \frac{1}{r^2} \frac{d}{dr} \bigg[ \rho v r^2 \bigg( \frac{1}{2} v^2 + \frac{\gamma}{\gamma -1} \frac{P}{\rho} \bigg) \bigg] = Q,  \label{eq:energy}
\end{align}
where the volumetric energy and mass injection rates are 
\begin{equation}
    q = \begin{cases}
    q_0 (R/r)^\Delta, & (r \leq R) \\ 
    0, & (r > R)  
    \end{cases}  \quad \mathrm{and} \quad
    Q = \begin{cases}
    Q_0 (R/r)^\Delta, & (r \leq R) \\ 
    0, & (r > R) 
    \end{cases} 
\end{equation}
respectively where $\Delta=0$ is taken for the CC85 model. Equations~\ref{eq:continuity}, \ref{eq:momentum}, and \ref{eq:energy} can be re-written as a single equation, the derivative of the Mach number, as
\begin{equation}
    \frac{d\mathcal{M}}{dr} = \frac{1}{(\gamma P / \rho )^{1/2}} \frac{dv}{dr} + \frac{v}{(\gamma P / \rho )^{1/2}} \frac{1}{2 \rho} \frac{d \rho}{dr} - \frac{v}{(\gamma P / \rho )^{1/2}} \frac{1}{2 P} \frac{dP}{dr}. 
    \label{eq:dMds}
\end{equation}
We then impose the boundary condition \citep{Chevalier1985,Wang1995}: 
\begin{align}
    & \mathcal{M}(r=R) = 1. \label{eq:boundaryout}
\end{align}
For $1 \neq \Delta < 3 $, the solution for the Mach number as a function of radius within the WDR is
\begin{align}
    \mathcal{M} & ^{1/(1-\Delta)}  \bigg[  \frac{2 + \mathcal{M}^2(\gamma -1) }{\gamma +1}  \bigg] ^{- \frac{1+\gamma}{2[(\Delta - 5) \gamma + \Delta - 1 ]}}  \nonumber \\
    & \times \bigg[ \frac{1- (\Delta - 3 ) \gamma \mathcal{M}^2 - \Delta}{1+3\gamma - (\gamma+1)\Delta} \bigg] ^{ \frac{(\Delta -3 ) \gamma + \Delta -1 }{(\Delta-1)[(\Delta -5) \gamma + \Delta -1 ]}} = \frac{r}{R}. \quad (r \leq R) 
    \label{eq:MachSolInterior}
\end{align}
For $\Delta =1$, the solution is
\begin{equation}
    e^{\frac{\mathcal{M}^2-1}{4 \mathcal{M}^2 \gamma}} \bigg( \gamma \mathcal{M}^2 \bigg)^{-\frac{\gamma+1}{8\gamma}} \bigg[ \frac{\gamma(2+\mathcal{M}^2(\gamma-1))}{\gamma+1} \bigg]^{\frac{1+1/\gamma}{8}}  = \frac{r}{R}. \quad (r \leq R) 
    \label{eq:MachSolInteriorDelta1}
\end{equation}
These solutions agree with those also derived by \citet{Zhang2017}. Taking $\Delta=0$, we arrive at the CC85 solution. The adiabatic, spherically expanding, exterior ($r>R$) solutions for the Mach number are identical to the CC85 solutions for all $\Delta$ models. 

Physical solutions require the definition of the total energy and mass injection rates, $\dot{E}_T$ and $\dot{M}_T$, within the WDR. We use 
\begin{equation}
    \dot{M}_T = \beta \ \dot{M}_\mathrm{SFR}, \quad  \mathrm{and} \quad \dot{E}_T = \alpha \ 3.1 \times 10^{41} \ \dot{M}_\mathrm{SFR,*} \ [\mathrm{ergs \ s^{-1}}],
    \label{eq:MdotEdotT}
\end{equation}
where $\alpha$ and $\beta$ are the dimensionless energy thermalization and mass-loading efficiencies, $\dot{M}_\mathrm{SFR,*} \equiv \dot{M}_\mathrm{SFR} / \mathrm{M_\odot \ yr^{-1}}$ is the dimensionless star-formation rate, and we have assumed that there is one supernova per 100\,$M_\odot$ of star formation and that each supernova releases $10^{51}$\,ergs of energy. To make a direct comparison with the uniform case $(\Delta=0)$, we normalize the rates to that of the uniform CC85 model. The normalization requirement for energy and mass loading, with rates $q' \propto Q' \propto r^{-\Delta} \ (\Delta < 3)$, are
\begin{align}
    q_0' = \frac{3-\Delta}{3} \frac{\dot{M}_T}{(4/3)\pi R^3} \ \ \mathrm{and} \  \ Q_0' = \frac{3-\Delta}{3} \frac{\dot{E}_T}{(4/3)\pi R^3 } \quad (\Delta<3)
    \label{eq:qQDotNorm}
\end{align}
From Equations~\ref{eq:continuity} and \ref{eq:energy}, the sound speed and velocity are  
{\footnotesize
\begin{equation}
    c_s^2 =   \frac{\dot{E}_T}{\dot{M}_T} \bigg[ \frac{(\gamma-1)\mathcal{M}^2 +2 }{2 (\gamma-1)} \bigg]^{-1} \ \mathrm{and} \ v = \mathcal{M} \ \bigg ( \frac{\dot{E}_T}{\dot{M}_T} \bigg)^{1/2} \bigg[ \frac{(\gamma-1)\mathcal{M}^2 +2}{2 (\gamma -1 ) } \bigg]^{-1/2}.
    \label{eq:soundspeed}
\end{equation}
}
The density is obtained from the continuity equation as 
\begin{equation}
    \rho = \frac{\dot{M}_T}{4\pi v} \frac{r^{1-\Delta}}{R^{3-\Delta}}, \ (r \leq R) \quad \mathrm{and} \quad \rho = \frac{\dot{M}_T}{4\pi v r^2},  \ (r > R ).
    \label{eq:density}
\end{equation}
The remaining quantity, the pressure, is solved from the sound speed as $P = \rho \, c_s^2 / \gamma $, where we take $\gamma = 5/3$ throughout the paper. 

\begin{figure*}
    \centering
    \includegraphics[width=\textwidth]{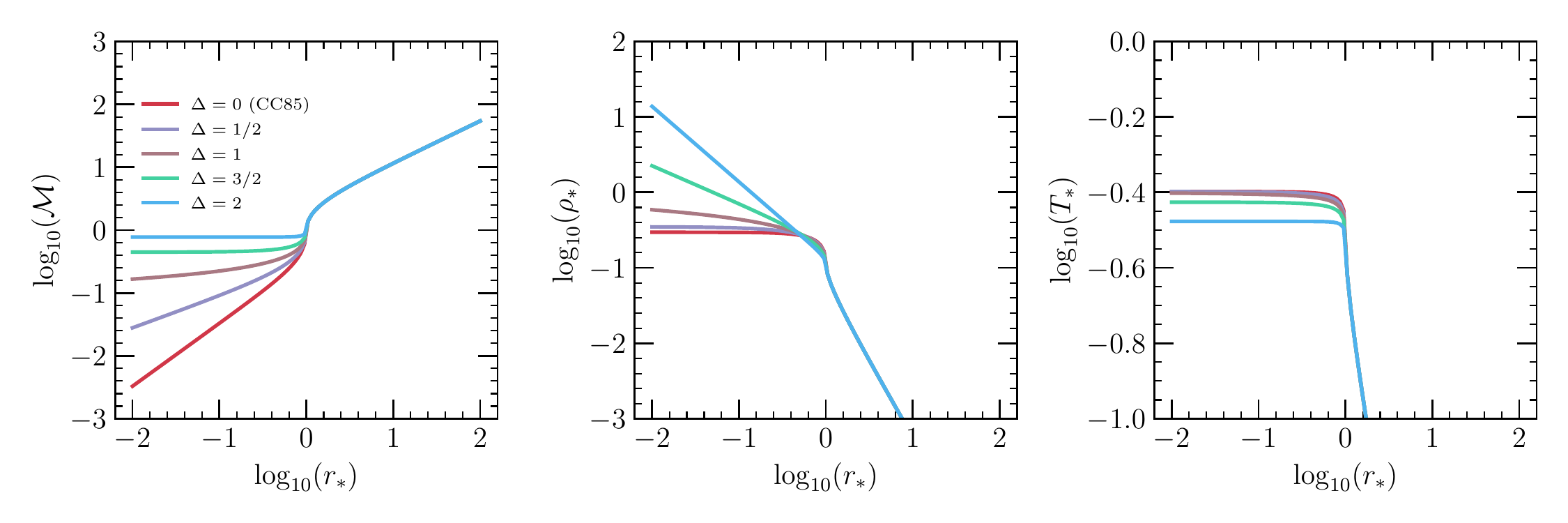}
    \caption{The Mach number, dimensionless density, and dimensionless temperatures for $\Delta=0\, (\mathrm{CC85}), \ 1/2, \ 1, \ 3/2, \ \mathrm{and} \ 2$ models as a function of the dimensionless radius $r_*=r/R$, assuming $\gamma=5/3$. For $r_* \ll 1$, $\mathcal{M} \rightarrow 0 $ for $\Delta=0$ and $\mathcal{M} = \sqrt{3/5} \simeq 0.77$ for $\Delta=2$ (see Eq.~\ref{eq:Mach_Constant}). The dimensionless solutions detail the structure of the outflow for arbitrary $\dot{E}_T$, $\dot{M}_T$, and $R$, see Table \ref{tab:centrals}.}
    \label{fig:dimensionlessdeltas}
\end{figure*}

In Figure~\ref{fig:dimensionlessdeltas}, we plot the dimensionless Mach number, density, and temperature. Relative to uniform injection ($\Delta=0$, red line), we find an isothermal sphere model ($\Delta=2$, blue line) produces a higher Mach number and denser outflow within the interior of the starburst. In Table~\ref{tab:centrals} we present analytic central limit solutions for $\Delta=0 \ \mathrm{and} \ 2$ models. We find the Mach number for an isothermal sphere ($\Delta=2$) is constant:
\begin{equation}
    \mathcal{M} = \sqrt{3/5} \simeq 0.77 \, \quad (\Delta=2, \ r<R).
    \label{eq:Mach_Constant} 
\end{equation} 
This starkly contrasts the Mach number for a CC85 (uniform) model, which linearly grows as $\mathcal{M} = 2^{-5/14} 3^{-11/14} r_* \simeq 0.33 r_*$ from the origin, where $r_* = r/R$. Both the pressure and density scale as $r^{-1}$ such that the velocity profile is also constant within the WDR: 
\begin{equation}
    v =   (\dot{E}_T / 3 \dot{M}_T)^{1/2},  \, \quad (\Delta=2, \ r<R).
    \label{eq:v_Constant_EdotMdot} 
\end{equation}
This can be written in terms of the asymptotic wind-velocity, \newline $v_\infty = (2 \  \dot{E}_T / \dot{M}_T )^{1/2}$, as 
\begin{equation}
    v = v_\infty / \sqrt{6} \simeq 0.4 \ v_\infty.     \, \quad (\Delta=2, \ r<R) 
    \label{eq:v_Constant_vasymp}
\end{equation}
We see that the interior flow for a $\Delta=2$ model is approximately half as fast as the energy-conserving supersonic terminal velocity. The difference in kinematics may be observable in velocity-resolved line profiles (see Sec.~\ref{sec:Obs_sig}). For $\Delta=1$, in the limit that $r_* \ll 1$, the Mach number is given by $\mathcal{M} = [-20/3 \times \ln((4/3)^{1/5}r_*)]^{1/2}$. The remaining quantities can be calculated by combining this with Equations~\ref{eq:soundspeed} and \ref{eq:density}. 
\begin{table}
 \caption{Analytic Central Solutions for $\gamma = 5/3$ for $\Delta=0$ and $\Delta=2$ models. See Section~\ref{sec:solutions} for central limit of the $\Delta=1$ model. The physical variables are written in cgs units and assume $\mu=0.6$.}
 \label{tab:centrals}
 \begin{tabular}{lcc}
 \hline
  \hline
  Dimensionless Variable & Uniform Sphere & Isothermal Sphere\\
                & $\Delta = 0$  & $\Delta = 2$ \\ 
                & $r_* \ll 1$   & $r_* \ll 1$ \\ 
  \hline
  $\mathcal{M}$ & $r_* / (2^{5/14} 3^{11/14}) $ & $\sqrt{3/5}$ \\ [0.1cm]
  $v/\dot{M}_T^{-1/2} \dot{E}_T^{1/2}$    & $2^{1/7} 3^{-9/7} r_*$    &  $1/\sqrt{3}$  \\[0.1cm]
  $T / \dot{M}_T^{-1} \dot{E}_T \mu m_p k_b^{-1}$ & $2/5$ & $1/3$ \\ [0.1cm]
  $\rho/\dot{M}_T^{3/2} \dot{E}_T^{-1/2} R^{-2} $ & $3^{9/7} / (4 \pi 2^{1/7})$    &  $\sqrt{3}/(4\pi r_*) $  \\[0.1cm]
  $P/\dot{M}_T^{1/2}\dot{E}_T^{1/2} R^{-2}$    & $3^{9/7} / (5 \pi 2^{8/7})$    &  $1/(4\pi \sqrt{3} r_*) $  \\
  \hline
  Physical Variable &  Uniform Sphere & Isothermal Sphere \\ 
  (solar metallicity, cgs units) & $\Delta = 0 $ & $\Delta = 2 $ \\ 
                     & $r_* \ll 1$   & $r_* \ll 1$ \\ 
  \hline
  $v/\alpha^{1/2}  \beta^{-1/2}$  & $1.885 \times 10^7 r_*$  & $4.048 \times 10^7 $ \hspace{0.4cm} \\
  $T/\alpha \ \beta^{-1}$  & $1.430 \times 10^7$ \hspace{0.15cm} & $1.192 \times 10^7$ \hspace{0.4cm} \\ 
  $n / \alpha^{-1/2}  \beta^{3/2} R^{-2} \dot{M}_\mathrm{SFR,*}$ & $2.649 \times 10^{41}$ \hspace{0.05cm} & $1.233 \times 10^{41}/r_*$  \\
  $P / \alpha^{1/2}  \beta^{1/2} R^{-2} \dot{M}_\mathrm{SFR,*}$  & $5.232 \times 10^{32}$ \hspace{0.05cm} & $2.030 \times 10^{32} / r_*$ \\
  \hline 
 \end{tabular}
\end{table}
Equations \ref{eq:MachSolInterior} and \ref{eq:MachSolInteriorDelta1} are solutions to an implicit equation. To use the solution, one is required to define an inner and outer radius. As shown in Table \ref{tab:centrals}, for an isothermal sphere model, there is a strong dependence on the inner radii, as the density and pressure diverge towards infinity. We define the inner radius (the minimum value of $r_*$) for the $\Delta=2$ model as $r_\mathrm{core,*}$. 

\subsection{Inference of the volumetric energy and mass injection rates within the wind-driving region}
A critical inference from X-ray observations of starburst nuclear centers are the energy thermalization and mass-loading efficiencies (i.e, $\alpha$ and $\beta$, see Eqs.~\ref{eq:MdotEdotT}). Using measurements of the central temperature and density, we infer $\alpha$ and $\beta$ \citep{Strickland2009} using the solutions from Table \ref{tab:centrals}, for both $\Delta=0$ and $\Delta=2$. These are:
\begin{align}
    \alpha(n,T)\simeq \begin{cases} 
    0.105\, n_{0.1}T_7^{3/2}R_{0.5}^2\dot{M}_{\rm 10}^{-1}, & (\Delta=0) \\
    0.297\, n_{0.1}T_7^{3/2}R_{0.5}^2\dot{M}_{\rm 10}^{-1} r_\mathrm{core,*}, &  (\Delta=2)
    \end{cases}
    \label{eq:alpha}
\end{align}
and
\begin{align}
    \beta(n,T) \simeq 
    \begin{cases}
    0.150 \, n_{0.1}T_7^{1/2}R_{0.5}^2\dot{M}_{\rm 10}^{-1}, & (\Delta=0) \\ 
    0.353 \, n_{0.1}T_7^{1/2}R_{0.5}^2\dot{M}_{\rm 10}^{-1}r_\mathrm{core,*}, & (\Delta=2)
    \end{cases}
    \label{eq:beta}
\end{align}
where $n_{0.1}=n/0.1$\,cm$^{-3}$, $T_7=T/10^7$\,K, $R_{0.5}=R/0.5$\,kpc, and $\dot{M}_{10}=\dot{M}_{\rm SFR,*}/10$.
From Equations~\ref{eq:alpha} and \ref{eq:beta}, it is apparent that inferred efficiencies $\alpha$ and $\beta$ from the  $\Delta=2$ model have a dependence on the defined inner radius $r_\mathrm{core,*}$, whereas for $\Delta=0$, there is not. The dependence on $r_\mathrm{core,*}$ for $\Delta=2$ arises from the diverging density and pressure profiles, see Table~\ref{tab:centrals}. 
\section{3D Hydrodynamic Simulations} 
\label{sec:cholla}
\begin{figure*}
    \centering
    \includegraphics[width=\textwidth]{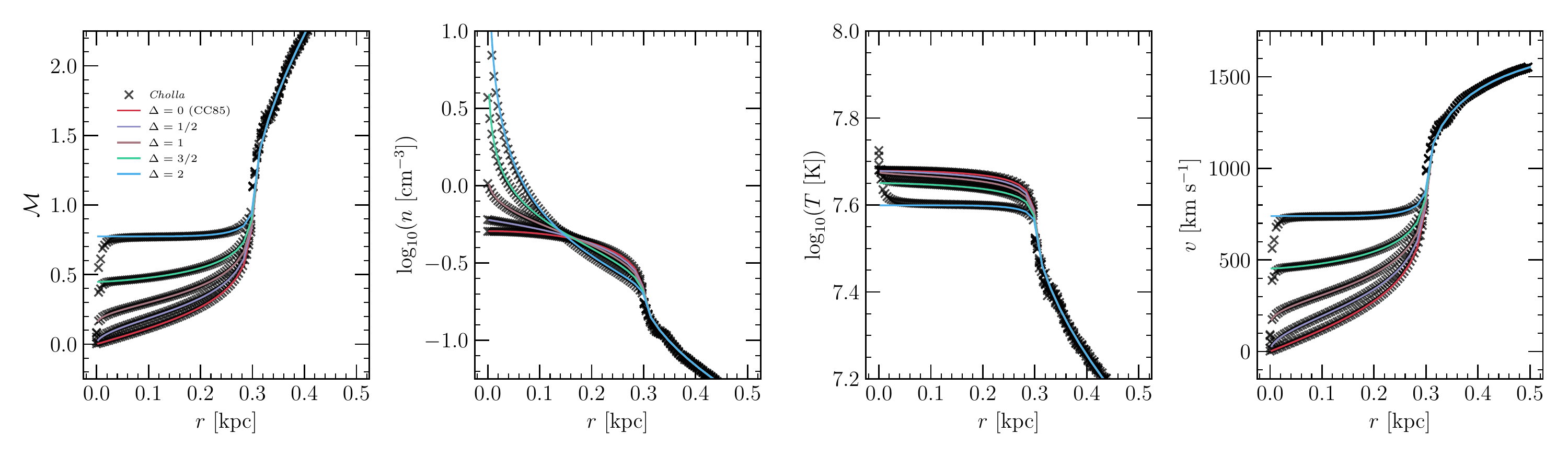}
    \caption{The Mach number, density, temperature, and velocity profiles from the \texttt{Cholla} simulations after steady-state has been reached (black x's), and analytic solutions (colored lines) for different injection slopes of $\Delta = 0, \ 1/2, \ 1, \ 3/2, \ \mathrm{and} \ 2$. All models assume the same total energy and mass injection rates $\dot{E}_T$ and $\dot{M}_T$ with $\alpha=1$, $\beta=0.3$, $R=0.3$ kpc, and $\dot{M}_\mathrm{SFR,*}=10$. The $\Delta=2$ flow maintains constant Mach $\mathcal{M}\simeq 0.77$ and $v \sim 750 \, \mathrm{km\,s^{-1}}$ for $r<R$ (Eqs.~\ref{eq:Mach_Constant} and \ref{eq:v_Constant_vasymp}).}
    \label{fig:compare}
\end{figure*}

We test our solutions using the  \texttt{Cholla} \citep{Schneider2015} code to simulate the starburst nuclei. The box has dimensions $1\,\mathrm{kpc^3}$ with $256^3$ cells, giving a cell resolution of $\Delta x \simeq 3.9$\,pc. 

Within a radius of $R$, we deposit energy and mass at a rate  $\dot{E}_T$ and $\dot{M}_T$ for different power-law injection slopes $\Delta = 0, \ 1/2, \ 1, \ 3/2, \ \mathrm{and} \ 2$, where the normalization for each $\Delta$ model is given by Equation~\ref{eq:qQDotNorm}. For all simulations, we take the M82-like fiducial wind parameters \citep{Strickland2009} of $\alpha =1 $, $\beta =0.3$, $R=0.3$\,kpc, and $\dot{M}_\mathrm{SFR,*} = 10$. The value of the core radius $r_\mathrm{*,core}$ is effectively set by the resolution. In order to make a direct comparison between the analytic solutions and numerical simulations, we do not include any additional physics, such as radiative cooling or gravity. For these wind model parameters, most of the flow is non-radiative and can escape a typical potential \citep{Chevalier1985,Thompson2016,Lochhaas2021}. All wind models reach a steady state, showing that the solutions are stable.

In Figure~\ref{fig:compare}, we show 1D radial profiles $+\hat{z}$ skewers of the Mach number, number density, temperature, and velocity profiles for both the analytic solutions (colored solid lines) and the \texttt{Cholla} simulations results (black x markers) after a time-steady solution has been established. The analytic solutions match the simulation results for every physical quantity. This implies that the imposed boundary condition of $\mathcal{M}=1$ at $r_*=1$, which was used in the analytic derivation, is indeed valid over the range of $\Delta$ values presented. Compared to the uniform sphere CC85 model ($\Delta=0)$, the isothermal sphere model ($\Delta=2$) maintains a much higher, constant, radial velocity $v \simeq 740$\,$\mathrm{km \ s^{-1}}$ throughout most of the WDR. 

\section{Observational Signatures}
\label{sec:Obs_sig}
\subsection{Surface Brightness}
\begin{figure}
    \centering
    \includegraphics[width=\columnwidth]{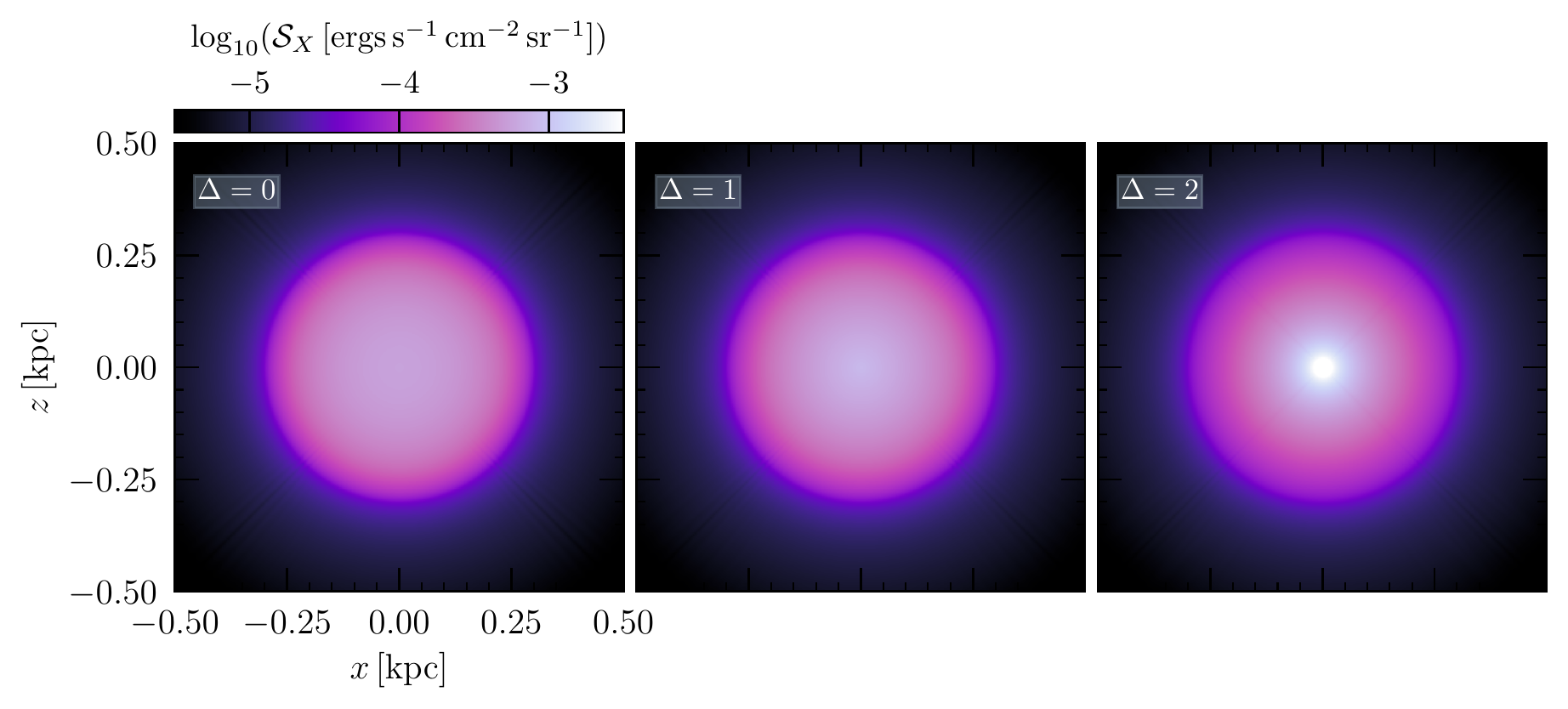}
    \caption{The X-ray surface brightness for $\Delta=0, \ 1, \ \mathrm{and} \ 2$ \texttt{Cholla} models at 1.5\,Myr with photons of energies in the observing band of XRISM ($0.3 \leq E \, [\mathrm{keV}] \leq 12$). The $\Delta=2$ model produces a strongly peaked surface X-ray brightness profile. }
    \label{fig:surface_bright}
\end{figure}

\begin{figure*}
    \centering
    \includegraphics[width=\textwidth]{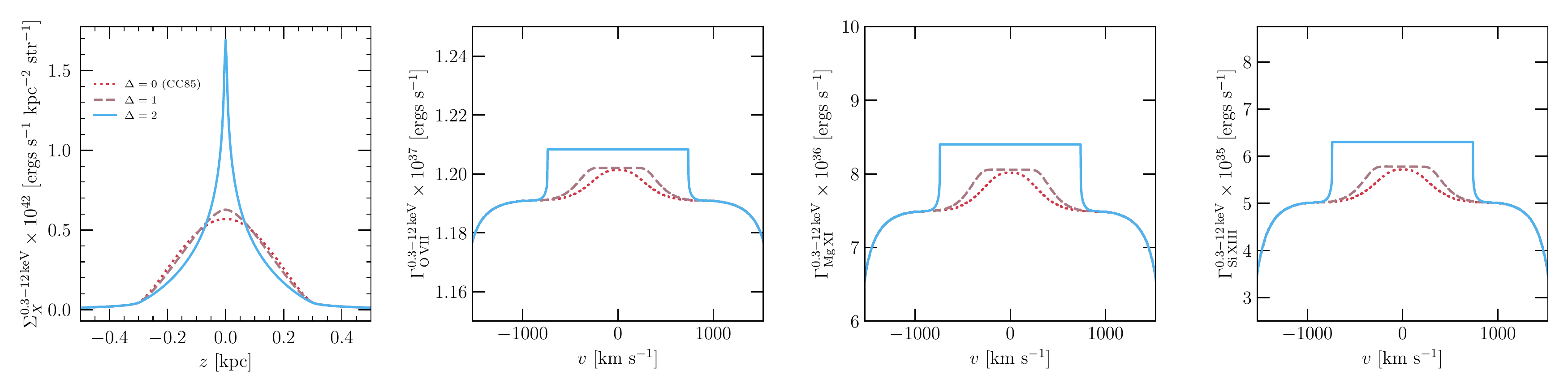}
    \caption{\textit{Left panel: }The X-ray surface brightness as a function of height for each $\Delta$ model with photons of energies $0.3 \leq E \, [\mathrm{keV}] \leq 12$. The $\Delta=2$ model produces a strongly peaked surface X-ray brightness profile, whereas the the CC85 X-ray surface brightness appears more broadened within the WDR. \textit{Right three panels: }The velocity resolved line profile with emissivities integrated over energies $0.3 \leq E\,\mathrm{[keV]} \leq 12$, for three He-like triplets: O {\sc vii}, Mg {\sc xi}, and Si {\sc xiii}, respectively. The emisivities are calculated with \texttt{PyAtomDB} with energies corresponding to the bandpass of XRISM's Resolve soft X-ray spectrometer instrument. We see that $\Delta=2$ models lead to a sharp horn-like feature in the velocity distribution across all He-like triplets, with the discrepancy between $\Delta=2$ and $\Delta=0$ more apparent in the heavier triplets.}
    \label{fig:bright_spec}
\end{figure*}

We calculate the instantaneous X-ray surface brightness as $\mathcal{S}^{\nu_1,\nu_2}_X(x,z) = \int_{\nu_1}^{\nu_2} d\nu \int_0^{L_y} dy \ n(x,y,z)^2 \Lambda(T(x,y,z),\nu),$ where $L_y$ is the length of the simulation domain, which includes the post-WDR supersonic wind. Using \texttt{PyAtomDB} \citep{2020pyatomdb}, we evaluate the plasma emissivity over XRISM's observing bandwidth ($0.3 \leq E_\gamma [\mathrm{keV}] \leq 12$), and assume solar metallicity abundances \citep{Anders1989}. In Figure~\ref{fig:surface_bright} we show $\mathcal{S}_X$ for $\Delta=0, \ 1, \ \mathrm{and} \ 2$ \texttt{Cholla} models. In the left panel of Figure~\ref{fig:bright_spec}, we calculate the surface brightness profile by taking $\Sigma_X$ and integrating along the $\hat{x}$ direction, and then dividing by the area of each surface. The surfaces are taken to be $50$\,pc$^2$. The $\Delta=2$ model leads to a strongly-peaked brightness profile, whereas the $\Delta=0$ model produces a less-peaked profile. We note that for the $\Delta=2$ model, the diverging density (see Tab.~\ref{tab:centrals}) implies a short cooling timescale. For these short cooling times, a cool non-X-ray emitting core may develop \citep{Wunsch2008,Lochhaas2021}. Radiative cores will be considered in a future work. 


\subsection{Velocity Resolved Line Profile}
XRISM's Resolve instrument is capable of resolving individual spectral lines and will trace gas motions through Doppler broadening and line shifts \citep{XRISM2020}. A spectrum of the entire wind-driving region will yield insight into the hot gas kinematics, which remain thus far unprobed. 
We construct resolved velocity line profiles for the $\Delta=0$ and $\Delta=2$ wind models. To do so, we consider shells inside $r\leq R$. When projected along the line of sight, this leads to a top-hat distribution in $n(r)$ versus $v(r)$ space, with bounds defined by $\pm v(r)$. We then calculate the emissivity of O {\sc vii}, Mg {\sc xi}, and Si {\sc xiii}, and integrate over XRISM's observing bandwidth. Next, we integrate over the volume $V = \sum_{\Delta r_i} 4 \pi r_i^2 \Delta r_i$. The result is shown in the three right panels of Figure~\ref{fig:bright_spec}. The $\Delta=2$ model has brighter emission along high velocities, whereas the $\Delta=0$ model is brightest where the gas is stationary. This is a result of the constant high velocity flow (Eq.~\ref{eq:v_Constant_vasymp}). For these injection parameters (see Sec.~\ref{sec:cholla}) the characteristic feature of the $\Delta=2$ model is the sharp increase in the emissivity at $v \sim \pm 750\,\mathrm{km\,s^{-1}}$. 
\section{Summary}
\label{sec:Summary}

In this work we study the dependence of injection slope, $\Delta$, for the kinematic and thermodynamic structure of the wind within the wind driving region. We derive analytic solutions, present their limits at small $r$ (see Table~\ref{tab:centrals}), and then confirm them with 3D \texttt{Cholla} simulations (see Fig.~\ref{fig:compare}). Importantly, we find that for a distribution of sources that scale as $r^{-2}$ ($\Delta=2$) the Mach number in the WDR is constant ($\mathcal{M} = \sqrt{3/5}$) and is approximately half of the asymptotic wind velocity (see Eq.~\ref{eq:v_Constant_vasymp}), faster than the uniform distribution \citep{Chevalier1985} or Schuster-like distributions \citep{Palou2013,Bustard2016}. The inferred energy and mass-loading efficiencies, $\alpha$ and $\beta$, are affected by $\Delta$, with a $\Delta=2$ sensitive to the core radius of injection. The $\Delta=2$ model produces strongly peaked X-ray brightness profiles (see Fig.~\ref{fig:surface_bright}). Figure \ref{fig:bright_spec} shows resolved line velocity profiles for relevant emission lines O {\sc vii}, Mg {\sc xi}, and Si {\sc xiii} for the $\Delta=0\,(\mathrm{CC85}),\ 1,\ \mathrm{and} \ 2 \ (\mathrm{isothermal \  sphere})$ models. These features may be observed by XRISM in the future. The $T$ and $n$ structure of the non-uniform models may be important in interpreting spatially-resolved maps for the interior of starburst superwinds. 

\citet{Wunsch2008,Lochhaas2021} showed that in cases of high mass-loading, the WDR develops a cool inert core. To make a direct comparison to the analytics, the simulations did not include cooling. We expect a cool core at the origin, as $\rho \propto r^{-1}$ ($\Delta=2$). This would affect the X-ray surface brightness profiles shown in Section \ref{sec:Obs_sig}. The condition for a cool inert core depends on the competing cooling and advection timescales and will be investigated in a future work. 

\section*{Acknowledgements} 
DDN and TAT thanks the OSU Galaxy Group and Chris Hirata for insightful conversations. DDN and TAT are supported by NSF\,\#1516967, NASA\,ATP\,80NSSC18K0526, and NASA\,21-ASTRO21-0174. E.E.S. acknowledges support from NASA\,TCAN\,80NSSC21K0271 and ATP\,80NSSC22K0720. SL and LAL were supported by NASA\,ADAP\,80NSSC22K0496. LAL is supported by the Simons Foundation, the Heising-Simons Foundation, and a Cottrell Scholar Award from the RCSA. 

\section*{Data Availability}
The data underlying this article will be shared on request to the corresponding author.

\appendix

\bibliographystyle{mnras}
\bibliography{bibliography} 


\bsp
\label{lastpage}
\end{document}